# Natural ventilation: a new method based on the Walton model applied to cross-ventilated buildings having two large external openings

Alain BASTIDE[1], Francis ALLARD[2] and Harry BOYER[1]
[1] Laboratoire de Physique du Bâtiment et des Systèmes,
40, Avenue de Soweto, 97410 Saint Pierre, Reunion Island, France
[2] Laboratoire d'Etude des Phénomènes de Transfert Appliqués,
Avenue Michel CREPEAU, 17042 La Rochelle Cedex 1, France

**Abstract**

In order to provide comfort in a low energy consumption building, it is preferable to use natural ventilation rather than HVAC systems. To achieve this, engineers need tools that predict the heat and mass transfers between the building's interior and exterior. This article presents a method implemented in some building software, and the results are compared to CFD. The results show that the knowledge model is not sufficiently well-described to identify all the physical phenomena and the relationships between them. A model is developed which introduces a new building-dependent coefficient allowing the use of Walton's model, as extended by Roldan to large external openings, and which better represents the turbulent phenomena near large external openings. The formulation of the mass flow rates is inversed to identify modeling problems. It appears that the discharge coefficient is not the only or best parameter to obtain an indoor static pressure compatible with CFD results, or to calculate more realistic mass flow rates.

## 1. Introduction

Natural ventilation is usually used during summer or hot seasons; during the cold season buildings are closed to preserve energy. The world energy context requires the prediction of both the seasonal heat production and the seasonal cold production over one year to give a reliable estimation of energy expenditure, and to limit the $CO_2$ production of active systems.

This article focuses on the mass transfers for natural ventilation and passive cooling during the summer season, by opening the building. To achieve a well-ventilated and comfortable building, architects ought to have tools that forecast and improve the energy consumption by a building, and the heat and mass transfers through large external openings. These tools have to integrate dynamic thermo-aeraulic predictions (Boyer *et al*, 1999) and various scenarios to cover the entire year in one simulation and to try and avoid the over-dimensioning of active air treatment systems.

Our methodology is to use the simulation results of detailed models and to adapt them to the discretization level generally in use in the thermal study of buildings, in order to calculate the mass flow rate and heat transfer between thermal zones. We chose the Walton model (Walton, 1984) modified by Roldan (Roldan, 1985 and Allard, 1992) for large external openings, and compared the numerical results with those produced by the $RNG-k-\varepsilon$ model (Yakhot, 1986). Walton proposes a methodology (Walton, 1989) for studying the problem of a building with an airflow network composed of infiltrations and large openings.

The Walton model (Walton, 1989) is attractive for the following reasons :

• The implementation is easy when the dynamic simulation software include cracks;

• The neutral axis is not required to calculate the mass transfer through the opening and so the computational program's algorithm is less complex but efficient;

• The number of constants to fix is reduced to one: the discharge coefficient for each kind of opening. The flow exponent is 0.5, which corresponds to turbulent air flow.

The Walton model gives a good prediction of gravitational flows through internal large openings in steady-state conditions (Bastide, 2004) compared to models based on neutral axis determination. Roldan (Roldan, 1985) proposed to modify this model for large openings placed on a facade. He added a term in the pressure calculation to take into account the external pressure, and to link it to the internal pressure.

The main difference between crack flow models and vertical large openings lies in the transformation of the kinetic energy into static pressure. In crack flow models, the mass flow rate is only governed by the pressure difference and the main assumption is that the kinetic energy is totally dissipated. For large opening models, streamlines cross the building. The important consequence of this is that a large part of the kinetic energy is transported through the building, without transferring kinetic energy into static pressure.

In this article, the model defined by Walton (Walton, 1984) and adapted by Roldan (Roldan, 1985) is improved to take into account the physics of the airflow that goes through a cross ventilated building. The resulting model predicts the mass flow that crosses the building with good accuracy relative to CFD predictions. The relative internal pressure of each model is compared and discussed.

## 2. Numerical investigation and modeling of the mass flow rate through large openings

### 2.1. Turbulence modeling

#### 2.1.1. Single zone building model

The computational domain is presented in Figure 1. All measurements are dimensionless. The building model is at the center of the ground's surface. The value of the parameter $L$ is 7 meters.

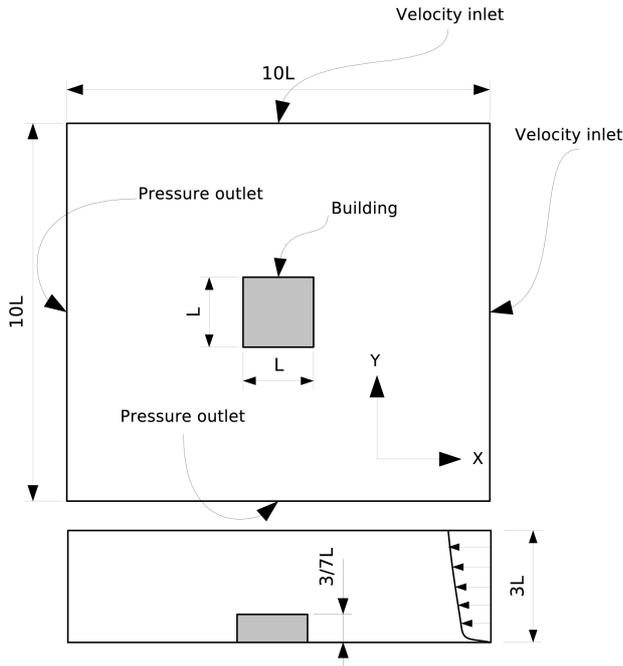

*Figure 1: Three-dimensional physical domain*

The building model shown in Figure 2 is cross-ventilated. Two opposite openings are located at height $H = 1\text{m}$. The areas of these two identical openings are $5.6\text{m}^2$.

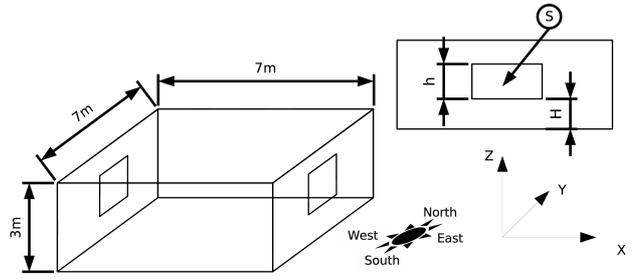

*Figure 2: The cross-ventilated building model and the orientation convention for each facade (East and West)*

#### 2.1.2. Introduction to CFD modeling

Evola and Popov (Evola and Popov 2006) compare the CFD predictions of four turbulence models with experimental observations. The numerical results obtained using the RNG method (Yakhot, 1986) show good agreement with the experimental data. The discrepancy between the calculated and expected ventilation rates is under 10%. The RNG model can be considered a useful tool for the study of airflow inside and around a building when dealing with wind-driven natural ventilation. The Large Eddy Simulations (LES) model may be preferable for advanced fluid dynamics that require higher precision.

#### 2.1.3. Turbulence models

The air is treated as an incompressible fluid with constant physical properties. The steady state equations of mass conservation and momentum are given below, applying the Reynolds decomposition in a tensorial form:

$$\frac{\partial U_j}{\partial x_j} = 0 \tag{1}$$

$$\frac{\partial}{\partial x_i}(U_i U_j) = -\frac{1}{\rho}\frac{\partial p}{\partial x_j} + \frac{\partial}{\partial x_i}\left[\nu\left(\frac{\partial U_j}{\partial x_i} + \frac{\partial U_i}{\partial x_j}\right) - \overline{u_i' u_j'}\right] \tag{2}$$

These equations include the term, $-\overline{u_i' u_j'}$ which represents the Reynolds stresses. This term models turbulent phenomena. The quality of the results depends on the choice of the class of model, and this depends on the set of equations used to solve the physical problem.

The Reynolds stress terms were modeled with the expression below, using the Boussinesq assumption of turbulent viscosity $\nu_t$. This approach assumes that the turbulent Reynolds stresses are linked to the set of averaged flow properties:

$$-\overline{u_i' u_j'} = \nu_t\left(\frac{\partial U_i}{\partial x_j} + \frac{\partial U_j}{\partial x_i}\right) - \frac{2}{3}k\delta_{ij} \tag{3}$$

$$\nu_t = C_\mu \frac{k^2}{\varepsilon} \tag{4}$$

These equations introduce two terms $k$ and $\varepsilon$ that are respectively the turbulent energy and the turbulent energy dissipation rate. Two closure equations have to be defined based on the transport of these quantities.

Several closure models have been developed in recent decades to close the system of equations resulting from the Reynolds decomposition, such as the $k-\varepsilon$ model (Launder and Spalding, 1974) and the Boussinesq assumption. However, the Boussinesq assumption used by many turbulence models is not always efficient. The concept of turbulent viscosity has been extended to try and model the anisotropy of turbulent flows. Quadratic and cubic models of turbulent viscosity have been developed. Bastide (Bastide 2004) established that this kind of model does not bring any significant improvement in flow predictions over a closed bluff body (Martinuzzi, 1993, Rodi, 1997 and Lakehal 1997) compared to the $RNG-k-\varepsilon$ model, but the results are better than the standard $k-\varepsilon$ model. Vortex and reattachment zones are either non-existent, under-predicted or over-predicted by the standard $k-\varepsilon$ model.

*2.1.4. The turbulence model of the renormalization group: $RNG-k-\varepsilon$*

The two closure equations are widely implemented in computational fluid dynamics software. The $RNG-k-\varepsilon$ model presents a double advantage compared to all the RANS models. The $RNG-k-\varepsilon$ model is designed to predict fluid flow for high and low Reynolds numbers, whereas the other RANS models are designed for either high or low Reynolds numbers. These two flow regimes are observed in an opened building, and so it appears necessary to apply a model such as $RNG-k-\varepsilon$.

The two closure equations of the $RNG-k-\varepsilon$ are given by the following expressions:

$$U_j \frac{\partial k}{\partial x_j} = \frac{\partial}{\partial x_j}\left[\left(\nu+\frac{\nu_t}{\sigma_k}\right)\frac{\partial k}{\partial x_j}\right]+P_k-\varepsilon \quad (5)$$

$$U_j \frac{\partial \varepsilon}{\partial x_j} = \frac{\partial}{\partial x_j}\left[\left(\nu+\frac{\nu_t}{\sigma_\varepsilon}\right)\frac{\partial \varepsilon}{\partial x_j}\right]+\frac{C_{\varepsilon 1}}{\varepsilon}k P_k - \frac{C_{\varepsilon 2}}{\varepsilon^2}k + R \quad (6)$$

The modification of the mathematic structure compared with the $k-\varepsilon$ model is located in the transport equation of the dissipation rate of turbulent kinetic energy $\varepsilon$. An new term $R$ is added to reduce the influence of the term $\frac{C_{\varepsilon 1}}{\varepsilon}k P_k$ and to trim down the turbulent kinetic energy production for $\eta > \eta_0$. The production term $P_k$ is calculated from gradient products.

$$P_k = \nu_t \left(\frac{\partial U_i}{\partial x_j}+\frac{\partial U_j}{\partial x_i}\right)\frac{\partial U_i}{\partial x_j} \quad (7)$$

$$R = \frac{C_\mu \eta^3 (1-\frac{\eta}{\eta_0})}{1+\chi \eta^3}\frac{\varepsilon^2}{k} \quad (8)$$

with $\eta = 2\left(\frac{P_k}{\nu_t}\right)^{\frac{1}{2}}\frac{k}{\varepsilon}$

The constants introduced in these equations have been calculated mathematically, unlike the traditional $k-\varepsilon$ models where they are determined from various experiments for different flow structures and regimes.

| $C_{\varepsilon 1}$ | $C_{\varepsilon 2}$ | $C_\mu$ | $\eta_0$ | $\chi$ | $\sigma_k$ | $\sigma_\varepsilon$ |
|---|---|---|---|---|---|---|
| 1.42 | 1.68 | 0.085 | 4.38 | 0.012 | 0.719 | 0.719 |

*Table 1: $RNG-k-\varepsilon$ model coefficients*

*2.1.5. Boundary conditions and wall treatment*

The boundary conditions are described in Figure 1. At the inlet, a log profile is set to model the atmospheric boundary layer. The value of the velocity reference at a height of 10m above the ground is $2.96\, m.s^{-1}$.

Some simulations which included a two layer turbulence model hardly improved the results for the mass flow rate, despite using much calculation time. The near wall cells are thus treated by a standard wall law using the assumption that the walls are smooth for the building model and slightly rough for the ground. The wall law uses the following coefficient: $E=9$. The roughness coefficient for the wall's log law is $z_0=0.039$.

*2.1.6. Grid independence test*

Three grids have been defined to verify that the observed results are grid-independent. Accurate results and a minimal number of cells were obtained at the end of the grid independence test. A grid of 411,000 cells was chosen to carry out all the numerical investigations.

**2.2. Surface Pressure calculations**

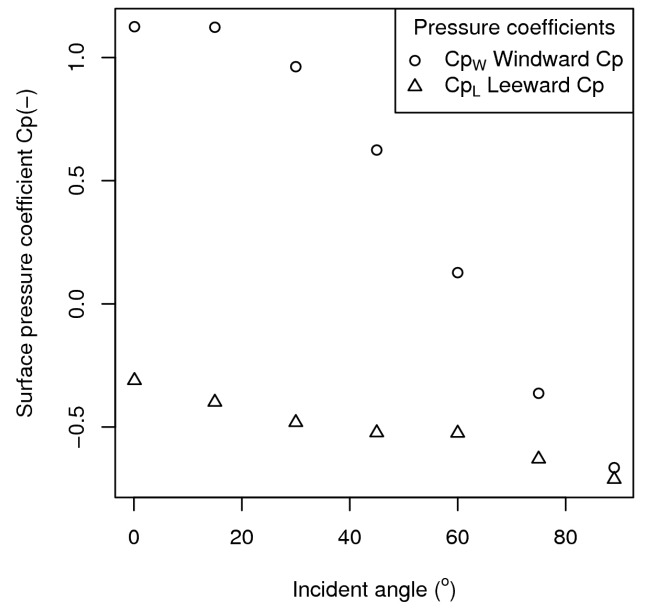

*Figure 3 : Windward and leeward surface pressure coefficients*

The global pressure coefficient for different incident angles and for a facade is evaluated from CFD simulations. The building model is closed and is shown in Figure 3.

The local surface pressure coefficients are calculated with the following equation :

$$C_p = mean\left(\frac{P_{surf}(z) - P_0}{P_{dyn}(z_0)}\right) \qquad (9)$$

$C_p$ is a dimensionless coefficient. It is the average surface pressure transformed into a relative pressure divided by the dynamic pressure predicted at height $z_0$. These coefficients are calculated from CFD predictions for a closed building. Figure 3 shows, as a function of some incident angles, the surface pressure coefficient evaluated for the windward and leeward facades excluding, in the calculation of the average, the edge pressures. The pressure difference between these two opposite facades is always positive. The mass flow will then be in a single direction.

## 2.3. Mass transfer through large openings

The mass transfer calculation is performed for the $N$ cells which are distributed in the enclosure of the building or for the area considered in the enclosure. The net mass flow at an opening is evaluated by : $\dot{m}_{opening} = \Sigma_i^N \dot{m}_i$. $\dot{m}_i$ is the mass flow rate that goes through one cell.

## 2.4. Mass transfer though internal large openings

The thermal behavior of buildings is a very complex problem. Different modes of heat and mass transfer appear in complex volumes and the interactions between different physical phenomena are significant. Detailed analysis, by CFD for example, of the airflow pattern through the cracks is incompatible with the computation of energy consumption over a heating and/or a cooling period. To study the global behavior of buildings, we need some simple and reliable models capable of correctly predicting indoor temperatures and air quality.

Various models of heat and mass transfer airflow through an internal building opening have been based on the Bernoulli equation : see Allard and Utsumi (Allard, 1992) or Walton (Walton, 1984). These models adopt the following principal assumptions : the flow is considered laminar, the flow is steady state, air is a non-viscous fluid, and turbulence and flux constrictions are represented respectively by two coefficients: the discharge coefficient and the airflow exponent.

Classical models developed to calculate heat and mass transfers through large openings find the location of the neutral axis, although the model developed by Walton does not use this methodology. This model can be considered as an intermediary method, somewhere between the crack flow models and the classical models based on finding the neutral axis to predict the mass transfer that goes through internal large openings.

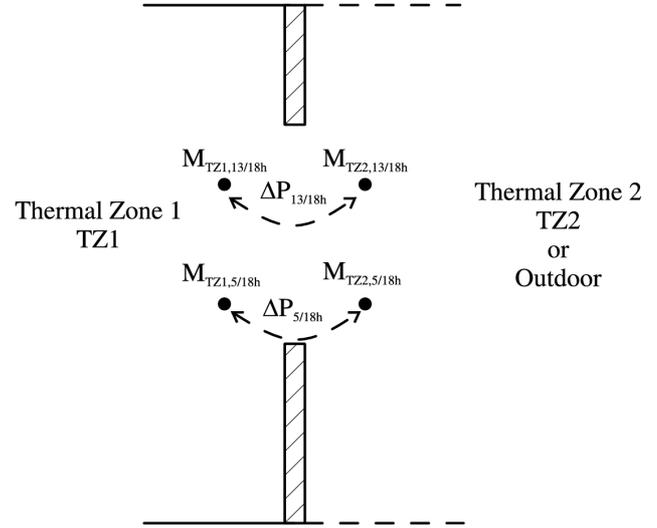

*Figure 4 : The locations of nodes ( M ) and pressure differences ( $\Delta P$ ) respectively in and between thermal zones*

Walton applied the methods of dimensional analysis $Nu/Pr = C\sqrt{Gr}/3$ to the heat and mass transfers through large openings. He concluded that the airflow through a large opening can be modeled by two small openings (Figure 4) placed at two special positions in a large opening. These positions are respectively located at 5/18 and 13/18 of the opening height. For non-zero thermal gradients between two thermal zones, the airflow between the top and the bottom of the openings is governed by the gradient of the hydrostatic pressures ( $\rho g z$ and $\rho = \rho_0 \frac{T_0}{T}$ ) at the openings. The assumption is that the density is uniform or linear in a thermal zone and the air density in a thermal zone is defined by its own air temperature.

In the presence of a density difference between two linked thermal zones, the airflow is bi-directional at an opening. The net mass flow rate is defined below:

$$\dot{m} = \dot{m}_{5/18h} + \dot{m}_{13/18h} \qquad (10)$$

For each small opening (i), the mass flow rate is given by the following expression:

$$\dot{m}_i = \rho C_{d,i} \frac{S}{2} sign(\Delta P_i)|\Delta P_i|^{1/2} \qquad (11)$$

The discharge coefficient in $C_d$ cluded in this definition represents two effects. The first effect is due to the friction between the air and the walls of the opening frame, and the intensity of this friction depends on the fluid viscosity. The second effect is due to the shape of the opening. The value of the discharge coefficient $C_d$ tends to one in the case of the opening area being almost equal to the wall area. This coefficient is valid for an air movement due to a thermal difference.

Walton obtained a discharge coefficient equal to for 0.78 an indoor opening separating two identical and closed thermal zones. Flourentzou (Flourentzou *et al*, 1998) evaluated a discharge coefficient close to 0.60 for an external large opening and for an *in situ* experiment in the absence of wind. In IEA Task 20 (Van der Mass, 1992), an experiment in a closed cell test composed of two thermal zones and a large opening between them showed that the average vertical discharge coefficients depend on the flow structure at the opening. The $C_d(z)$ vary between 0.0 and 0.66. With regard to this point, the $C_d(z)$ have been observed to not have locally constant values at the opening (Karava1, 2004). The value of the discharge coefficient is subject to a large degree of uncertainty. However, we need to define a set of constants that are compatible with the knowledge model and with the space dimension. In the kind of model described here, an average value of the discharge coefficient has to be fixed and adjusted on a case-by-case basis to well-represent the mass flow rate at an opening.

**2.5. Application of the Walton model to large external openings**

Roldan (Roldan, 1985 and Allard, 1992) proposed to apply the Walton method to large external openings. In the absence of wind, this modeling applied to an external large opening is coherent with the assumptions made by Walton. However, if the wind is present within the building, the pressure field changes around and in the building, and so Roldan added a pressure term in the pressure difference to take account of the external pressure field to predict the pressure in the building. He proposed a modification to the discharge coefficient included in this relation, in order to adjust the model to observations or to numerical predictions.

The Walton model of bi-directional airflow across an opening is given by the following expression:

$$\dot{m} = \rho C_d \frac{S}{2} sign\left(\Delta P_{\frac{5}{18}h}\right)\left|\Delta P_{\frac{5}{18}h}\right|^{1/2} + \quad (12)$$

$$\rho C_d \frac{S}{2} sign\left(\Delta P_{\frac{13}{18}h}\right)\left|\Delta P_{\frac{13}{18}h}\right|^{1/2} \quad (13)$$

The pressure gradients are for two thermal zones *ZTH1* and *ZTH2* :

$$\Delta P_{\frac{5}{18}h} = P_{ZTH1} - P_{ZTH2} - g\left(H + h\frac{5}{18}\right)\left[\rho_{ZTH1} - \rho_{ZTH2}\right] \quad (14)$$

$$\Delta P_{\frac{13}{18}h} = P_{ZTH1} - P_{ZTH2} - g\left(H + h\frac{13}{18}\right)\left[\rho_{ZTH1} - \rho_{ZTH2}\right] \quad (15)$$

he unknown factors of the problem are the pressure $P_{ZTH1}$ and $P_{ZTH2}$ calculated by a iterative algorithm like Newton-Raphson method.

In this article, this model is compared to CFD predictions and a new approach is developed to optimize the Roldan model.

*2.5.1. Influence of the external pressure distribution on mass flow rate*

The pressure difference between the leeward and windward facades depends on the geometrical aspects of the building, the atmospheric boundary layer profile, the incident angle and the intensity of the reference velocity. The velocity and pressure field around the building change spatially. The pressure field around a building is generally characterized by regions of over-pressure on the windward facades, and under-pressure on the facades parallel to the air stream and on the leeward side. When the building is opened, the pressure difference induces a mass flow through the building. This mass flow is dependent on the pressure difference field around the building.

*2.5.2. The streamline deformation near the building*

Ohba *et al* (Ohba, 2001) give a correction to the incident angle. The resulting angle incorporates the deformation of streamline tubes due to the upstream airflow complexity. The form of the streamline tubes close to the external opening is investigated by Hul *et al* (Hul, 2005). The streamlines do not follow the forward incident angle close to the wall, but depend on the form of the building, the shape and the configuration of the building openings and the incident angle. The airflow direction through the windward opening changes and the effective area 'seen' by the streamline tubes changes too. Ohba *et al* (Ohba, 2001) propose correcting the incident angle of the wind using the fraction $\frac{A}{A_0}$ where $A = \frac{1}{\rho} \frac{\dot{m}}{V_p}$, $A = \frac{Q}{V}$ and wher $V_p$ e is the penetrating velocity and $\dot{m}$ the mass flow rate. The key point of this method exposed by Ohba *et al* (Ohba, 2001) is to measure the velocity $V_p$ at the central position close to the inlet of the building and to suppose that this velocity represents the mean velocity at the inlet. Using numerical fluid dynamics, the effective area can be calculated from the discretization of the volume opening. The effective area is given by the mean velocity : $\bar{V} = \frac{1}{N} \sum_i U_i$. The new effective area is: $A = \sum_i S_i U_i / \frac{1}{N} \sum_i U_i$.

The new expression for the effective area greatly increases the number of velocities and the accuracy of the mean velocity and the mass flow rate, which improves the precision of the mean velocity value. But the results are too far from the results obtained by using the penetrating velocity located at the center of opening. This methodology based on the streamline pattern has not been included in the model of mass flow rate formulated in this article because the discharge coefficient is considered equal or practically equal for large openings and a new term is include in our system of equations.

*2.5.3. Influence of the CFD modeling on predicted pressure distribution and mass flow rate predictions*

Results obtained by numerical simulations have to be accurate for the two physical quantities $\dot{m}$ and $P$. The mass flow and pressure differences are observed and used to define new knowledge models.

The turbulence modeling influences the quality of the CFD predictions. Evola and Popov (Evola and Popov 2006) show that the predictions and empirical values are close to the mean ventilation rate through a cross-building. The RNG model predicts the mass flow rate with an error lower than 10%, whereas the standard model over-predicts the mass flow rate (by more than 15%). The LES model has a relative error of 6.5% compared to experiments. The expected mean pressure coefficients are under 10% for the windward and leeward faces. The $RNG-k-\varepsilon$ model seems to be a good compromise with regard to the computational time and the quality of the predictions for this kind of problem.

Hu *et al* show that the structure of ventilating flow is better represented with the $SST-k-\omega$ model than with LES (Wilcox, 1998 and Menter, 1994). The pressure distribution on faces, the pressure extrema and the characteristic lengths are analyzed by Endo *et al* (Endo, 2005). They conclude that the $SST-k-\omega$ model is not accurate for these three quantities. $RNG-k-\varepsilon$ is a good compromise between the RANS models for predicting mass flow rate and pressure distribution.

The mass flow rates through the building are shown in Figure 3. Three RANS turbulence models have been employed : $k-\varepsilon$, $RNG-k-\varepsilon$ and $SST-k-\omega$.

The mass flow rates for an incident angle in the range $[30-90]$ are similar for the two models $k-\varepsilon$ and $RNG-k-\varepsilon$. Elsewhere, the $k-\varepsilon$ results have better residuals than those from the $RNG-k-\varepsilon$ modeling. The $k-\varepsilon$ standard model gives good agreement, but as Endo et al (Endo, 2005) found, the predicted pressure distribution is poor. The mass flow is over-predicted, and the pressure distribution poorly placed by the $SST-k-\omega$ model. These results demonstrate that the $RNG-k-\varepsilon$ model seems to be effective in the specific case of mass flow modeling.

*2.5.4. The model proposed by Roldan : an adaptation of the Walton model for external large openings*

Roldan (Roldan, 1985) proposed a modification of the mathematical formulation of the pressure difference included in the Walton model. He included the static wind-induced pressure on a facade in the pressure difference.

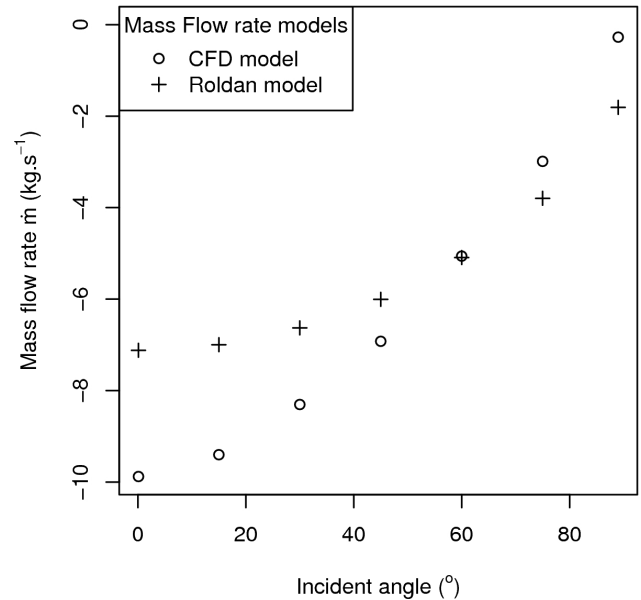

*Figure 6 : Mass flow rate calculated by the detailed model (CFD) and by the model developed by Roldan*

The pressure difference links the internal pressure and the external pressure on facade. Roldan suggested adjusting the discharge coefficient too. Preserving the mass flow model of equation 11, the mass balance of the thermal zone is calculated assuming that each pressure model for each node is as given in Table 2.

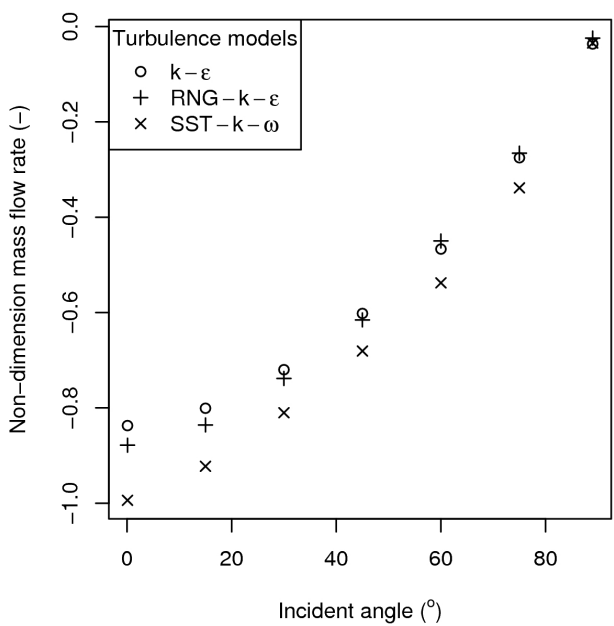

*Figure 5 : Mass flow rate through the windward opening predicted from three turbulence models.*

| Nodes | Roldan modifications |
|---|---|
| $M_{\frac{5}{18}h, TZI}$ | $P_{MTZI} = P_1 - \rho g \left[ H + \frac{5}{18} h \right]$ |
| $M_{\frac{5}{18}h, out}$ | $P_{Mout} = P_{out} - \rho g \left[ H + \frac{5}{18} h \right] + \frac{1}{2} \rho C_p V_{out}^2 \left( H + \frac{5}{18} h \right)$ |
| $M_{\frac{13}{18}h, TZI}$ | $P_{MTZI} = P_1 - \rho g \left[ H + \frac{13}{18} h \right]$ |
| $M_{\frac{13}{18}h, out}$ | $P_{Mout} = P_{out} - \rho g \left[ H + \frac{13}{18} h \right] + \frac{1}{2} \rho C_p V_{out}^2 \left( H + \frac{13}{18} h \right)$ |

*Table 2: nodes and their corresponding models in Roldan's model, after the Walton model.*

The model of pressure nodes mentioned above does not include the fact that a part of the kinetic energy goes through the building without transferring a fraction of its energy into kinematic pressure. In the unknown pressure $P_1$, a static pressure and a dynamic pressure is calculated. But we do not know the fraction of this dynamic pressure that is converted into kinematic pressure or static pressure. Roldan (Roldan, 1985) proposed adjusting the discharge coefficient to improve the model, but this coefficient is generally used to take turbulence effects and flow constriction into account. An adaptation of this model is thus needed.

In Figure 6, we observe that the results obtained by the Roldan (Roldan, 1985) model are either 30% lower, or twice as large as the values that result calculated using the CFD model. The two sets of points are not homothetic, and so this problem cannot be settled by adjusting a single discharge coefficient.

## 2.6. Inclusion of the predicted physical parameters to modify and improve the Roldan model

### 2.6.1. Introduction

Airflow through large openings involves a number of different physical phenomena, including steady-state gravitational flows, fluctuating flows resulting from wind turbulence, and re-circulation flows caused by boundary layer effects. The complexity of this problem has to be included in the knowledge model. The Roldan (Roldan, 1985) model adds to the Walton (Walton, 1984) model the external pressure differences but the prediction of mass flow rate does not reproduce the CFD predictions. The model presented by Roldan (Roldan, 1985) is therefore modified to take all the effects mentioned into account. An artificial pressure is evaluated using the inversion of the knowledge model and included to predict the mass flow rate and the relative pressure in the building model.

### 2.6.2. Formulation of an artificial pressure difference by applying the inverse problem

The pressure difference $\Delta P$ calculated from the external wind and external static pressure does not produce the expected predictions (Figures 7 and 8). The model is not sufficiently detailed to suitably predict the mass exchanges through large external openings. We propose to inverse the problem and thus to create an artificial pressure difference $\delta P_{CFD}$ in order to identify the modeling problems in Roldan's model.

The mass flow rate $\dot{m}$ is given by CFD predictions. The inverse pressure difference is given by the equation:

$$\delta P_{CFD} = sign(\dot{m}_{CFD}) \left( \frac{2 \dot{m}_{CFD}}{C_d \rho S} \right)^2 \qquad (16)$$

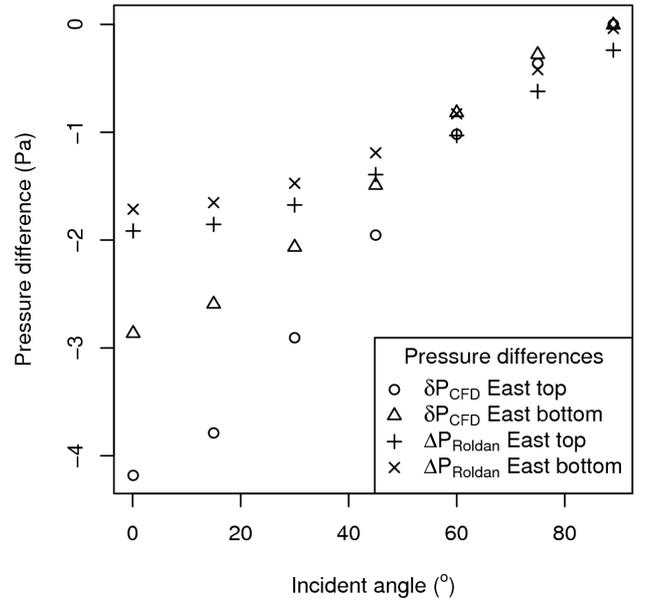

*Figure 7 : East opening, pressure differences determined following the inverse problem for the two small openings using the Walton model and the pressure differences calculated with the Roldan model*

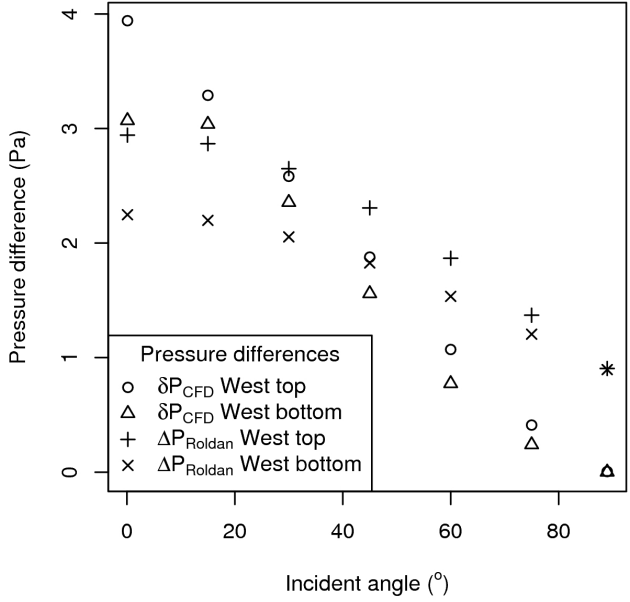

*Figure 8 : West opening, pressure differences determined following the inverse problem for the two small openings using the Walton model and the pressure differences calculated with the Roldan model*

$\delta P_{CFD}$ results from a mathematical construction that depends on the mass flow rate predicted by the CFD model and depends on the discharge coefficient determined for a gravitational flow without static pressure produced by the wind.

The slope formed (Figure 7) by the values of $\delta P_{CFD, East, top}$ is steeper than the values of $\delta P_{CFD, East, bottom}$. This difference is due to the shape of the atmospheric boundary layer and to a three-dimensional recirculation present above the ground on the windward facade. The airflow thus passes (Figure 7 and 8) mainly through the opening at the top. The same differences are observed for $\Delta P_{Roldan}$ but the pressure differences are not significant for an incident angle lying between $0°$ and $45°$.

In the range $[0°, 45°]$ (Figure 7) the results are similar because the airflow pattern is modified by the building model. Below $45°$, the streamlines follow the building facade. Airflow enters the building with a specific shape governed mainly by the pressure differences due to the air motion, which is orthogonal to the natural air displacement through a windward external large opening.

In Figure 8, between $20°$ and $90°$ $\delta P_{CFD}$ the are practically linear. For $90°$, the pressure differences are close to zero. There is no difference in behavior between the top and bottom of the leeward opening because the motion is mainly due to static pressure differences between the interior and the exterior. The situation here *could be* modeled as an airflow through a large internal opening.

However, the values determined by Roldan's model follow a parabolic shape (Figure 8). This is caused by the predicted indoor pressure in this model that includes a dynamic pressure.

### 2.6.3. Modeling of the building pressure coefficient $C_B$ for each small opening in the Walton sense

Thanks to the pressure difference $\delta P_{CFD}$ calculated above, a building pressure coefficient $C_B$ is defined. This coefficient models the part of kinetic energy that crosses one *small opening* of an external large opening. The mathematical formulation, described below, results from allowing for the surface pressure of the facade ($C_P$) and the mass flow rate through $\delta P_{CFD}$.

$$C_{B, top} = \frac{\delta P_{CFD, top}}{\frac{1}{2}\rho V_{out}^2 \left(H + \frac{13}{18}h\right)} + C_p \quad (17)$$

$$C_{B, bottom} = \frac{\delta P_{CFD, bottom}}{\frac{1}{2}\rho V_{out}^2 \left(H + \frac{5}{18}h\right)} + C_p \quad (18)$$

These non-dimensional coefficients are incorporated into the model developed by Roldan (Roldan, 1985) to constitute a new model (Table 3) to predict the mass flow rate through cross building models. This model can be easily included in classical energy building software like the crack models.

| Nodes | New formulation |
|---|---|
| $M_{\frac{5}{18}h, TZI}$ | $P_{MTZI} = P_1 - \rho g\left[H + \frac{5}{18}h\right] + \frac{1}{2}\rho C_{B, top} V_{out}^2 \left(H + \frac{5}{18}h\right)$ |
| $M_{\frac{5}{18}h, out}$ | $P_{Mout} = P_1 - \rho g\left[H + \frac{5}{18}h\right] + \frac{1}{2}\rho C_p V_{out}^2 \left(H + \frac{5}{18}h\right)$ |
| $M_{\frac{13}{18}h, TZI}$ | $P_{MTZI} = P_1 - \rho g\left[H + \frac{13}{18}h\right] + \frac{1}{2}\rho C_{B, bottom} V_{out}^2 \left(H + \frac{13}{18}h\right)$ |
| $M_{\frac{13}{18}h, out}$ | $P_{Mout} = P_1 - \rho g\left[H + \frac{13}{18}h\right] + \frac{1}{2}\rho C_p V_{out}^2 \left(H + \frac{13}{18}h\right)$ |

*Table 3: The new formalisation of the pressures including the pressures deduced with the inverse problem*

Equations 17 and 18 give, respectively, expressions for the building coefficient of the opening placed at the top and at the bottom of the opening.

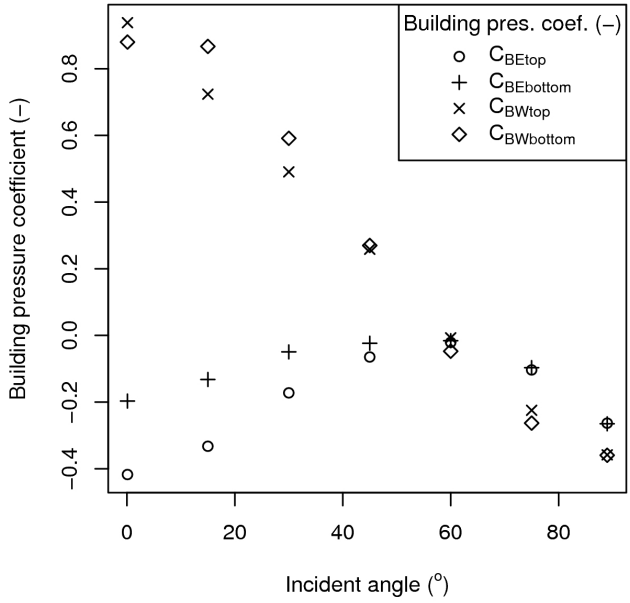

*Figure 9 : Building coefficients $C_B$ for each small opening in the Walton sense*

The values of the mass flow rates for each boundary condition of the numerical experiments and for each small opening of these coefficients are plotted in Figure 9. Following equations 17 and 18, the building pressure coefficients act on the pressure difference at openings. For the East external opening and the small opening placed on top, we notice that the building pressure coefficients are much higher than the surface pressure on the leeward facade. These coefficients are thus present to increase the mass flow through an external large opening.

The resulting model and these outputs (Figure 10) show that the improvement of the knowledge model developed by Roldan (Roldan, 1985) have allowed the physical phenomena to be better represented. The mass flow rate deduced by this new formulation of pressure differences is close to the numerical experiments.

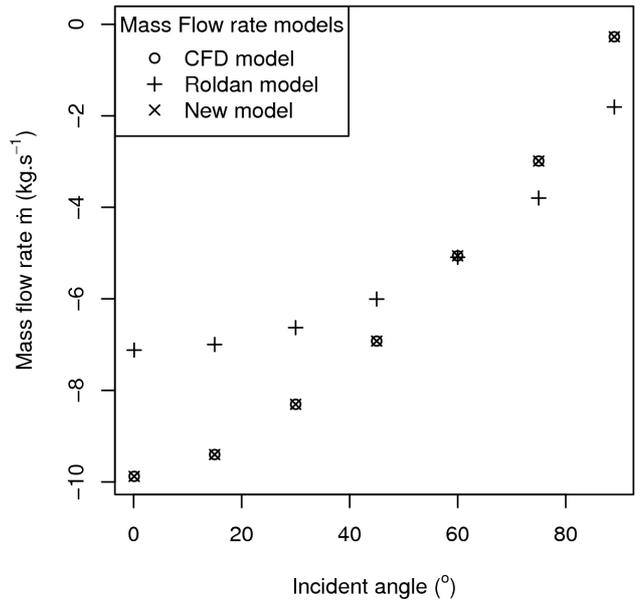

*Figure 10: Mass flow rates calculated for each knowledge model described above*

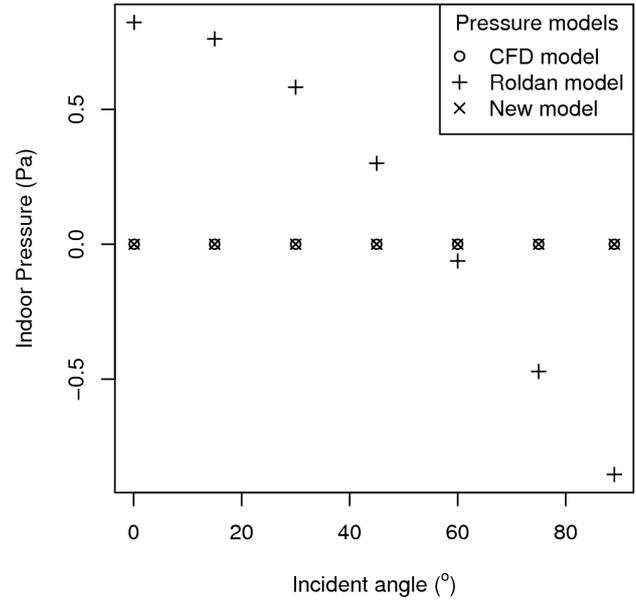

*Figure 11: Indoor kinematic pressure calculated for each model*

Figure 11 shows the pressure calculated for the models detailed in this article. We note that the kinematic pressure calculated by Roldan's model (Roldan, 1985) differs greatly from the numerical experimental results. The shape of the curve and the model detailed in Table 1 show that the pressure determined by this model includes a dynamic pressure. This dynamic pressure skews the problem because the pressure P defined by Walton (Walton, 1984), which partially crosses the building is static, and not dynamic. The pressure model developed by Roldan cannot be implemented to evaluate pressures in buildings.

A major implication is that if the Roldan model is employed to analyze heat and mass transfers in a complex building or multizone building then heat transfers, mass transfers and thermal comfort will be poorly predicted. On the other hand, in the absence of wind the Walton model will correctly calculate heat and mass transfers.

## 3. Conclusion

Mass flow rate predictions through large external openings deduced from the Walton-Roldan model differ greatly from CFD results. The knowledge model developed by Roldan proposes to include the surface pressure induced by wind in the pressure difference between the interior and the exterior. He explains that it is then necessary to set the discharge coefficient for each incident angle. However, the indoor pressure calculated by this method includes a dynamic pressure that corresponds to the kinetic energy that crosses the building without transferring all this former energy into kinematic pressure. The problem could not be understood as a crack-flow.

Building coefficients are then introduced to take into account this kinetic energy that passes through, and to improve the model implemented by Roldan. Energy present in the stream tube developed at the windward opening can now partly go through the building and transfer a part of its energy into potential energy through kinematic pressure.

Results of these modifications show that indoor pressures are close to the numerical experimental results, mass flows are better predicted and discharge coefficients are not changed. The discharge coefficient is kept to the value defined for a modeling problem without wind.

The indoor pressure evaluated by solving the mass balance is now kinematic, and not a dynamic pressure plus a static pressure. This model can now be coupled to a crack-flow model that needs a static pressure.